\documentclass[a4paper]{jpconf}
\usepackage{graphicx}
\begin{document}
\title{Continuum reverberation mapping in a $z$ = 1.41 radio-loud quasar}

\author{L J Goicoechea, V N Shalyapin, R Gil-Merino and V F Braga}

\address{GLENDAMA Project Team, Facultad de Ciencias, Universidad de Cantabria, Avda. 
de Los Castros s/n, 39005 Santander, Spain}

\ead{goicol@unican.es, vshal@ukr.net, r.gilmerino@gmail.com, bravi4@hotmail.com}

\begin{abstract}
Q0957+561 was the first discovered gravitationally lensed quasar. The mirage shows two images of 
a radio-loud quasar at redshift $z$ = 1.41. The time lag between these two images is well 
established around one year. We detected a very prominent variation in the optical brightness of 
Q0957+561A at the beginning of 2009, which allowed us to predict the presence of significant 
intrinsic variations in multi-wavelength light curves of Q0957+561B over the first semester of 
2010. To study the predicted brightness fluctuations of Q0957+561B, we conducted an X-ray, NUV, 
optical and NIR monitoring campaign using both ground-based and space-based facilities. The 
continuum NUV-optical light curves revealed evidence of a centrally irradiated, standard 
accretion disk. In this paper, we focus on the radial structure of the standard accretion disk 
and the nature of the central irradiating source in the distant radio-loud active galactic 
nucleus (AGN).
\end{abstract}

\section{Introduction}
Reverberation (or echo) mapping is a time-domain technique to resolve the accretion flow in 
AGNs. This relies on the analysis of time-delayed responses of different emitting regions to 
original fluctuations in an irradiating source [1]. While concurrent X-ray-UV-optical continuum 
monitoring campaigns of low-redshift AGNs are leading to puzzling results (e.g., [2]), there is 
only one AGN at $z \geq$ 1 with a continuum reverberation map: Q0957+561 [3]. The $z$ = 1.41 
radio-loud quasar Q0957+561 suffers a strong gravitational lens effect, so one observes two 
images Q0957+561A and Q0957+561B from Earth. Intrinsic flux variations in Q0957+561B lag those 
in Q0957+561A by about 14 months [4], and we taken advantage of this time delay between images. 
Optical follow-up observations in late 2008 and the first semester of 2009 showed very prominent 
variations in Q0957+561A [5], which enabled us to know in advance the flaring behaviour of 
Q0957+561B during the first semester of 2010, and to use the Liverpool Robotic Telescope (Sloan 
$griz$ bands), the UVOT on board the Swift satellite ($U$ band) and the Chandra Space Telescope 
(X-rays) for such a time period.

From the $Ugr$ light curves of Q0957+561B in 2010, we measured interband delays of several days 
that were consistent with a centrally irradiated, standard accretion disk [3]. In this scenario, 
a flaring source very close to the central supermassive black hole illuminates a standard 
accretion disk. The central flares are then thermally reprocessed into FUV-MUV variations in the 
inner disk to produce the observed NUV-optical variability. In Section 2 we present the NIR 
observations and the corresponding light curves, which were not discussed in [3]. These new NIR 
data are used to check the accretion scenario and discuss in detail the disk radial structure. 
In Section 3 we infer possible reconstructions of the effective luminosity of the central 
irradiating source from the NUV-optical-NIR light curves. Our main conclusions appear in Section 
4.  

\section{NIR observations and the accretion scenario}
In 2010, we conducted a NIR monitoring programme of Q0957+561 with the RATCam CCD camera on the 
Liverpool Robotic Telescope at La Palma, Canary Islands. The exposure times were 120 and 240 s 
per night in the $i$ and $z$ bands, respectively. The pre-processing steps included in the 
telescope pipeline are: bias subtraction, overscan trimming and flatfielding. In addition, we 
interpolate over bad pixels using the bad pixel mask, clean some cosmic rays and remove fringe 
patterns. Our crowded-field photometry pipeline is then used to produce instrumental fluxes of 
both quasar images over 54 epochs. To achieve a reasonable compromise between photometric 
quality and sampling rate, later we apply thresholds on the signal-to-noise ratio in the $i$ and 
$z$ bands. The final processing stage consists of the conversion from instrumental magnitudes to 
physical fluxes (in mJy), and the final NIR fluxes are available at 
\verb"http://grupos.unican.es/glendama/LQLMII_DR.htm". We also display the NIR light curves of 
Q0957+561B in Fig. 1a. This includes 50 data points at $\lambda_{\rm o}$ = 7481 \AA\ ($i$-band) 
and 45 data points at $\lambda_{\rm o}$ = 8931 \AA\ ($z$-band). While the $i$-band flux accuracy 
is 1.4\%, the $z$-band flux accuracy is only 2.0\%. Fig. 1b includes all NUV-optical-NIR records
of Q0957+561B in 2010.  

\begin{figure}[h]
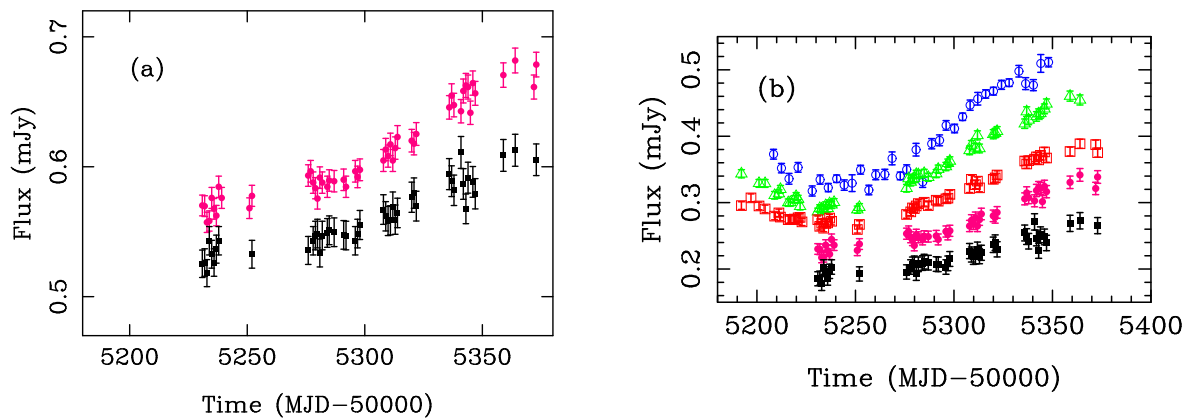

\begin{minipage}{17pc}
\includegraphics[angle=-90,width=17pc]{Fig1a.eps}
\end{minipage}
\hspace{2pc}
\begin{minipage}{17pc}
\includegraphics[angle=-90,width=17pc]{Fig1b.eps}
\end{minipage}
\caption{\label{Fig1}(a) Fluxes of Q0957+561B in the $i$ [\fullcircle] and $z$ [\fullsquare] 
bands of the SDSS photometric system in 2010. (b) NUV-optical-NIR light curves of Q0957+561B in 
2010: $U$ [\opencircle], $g$ - 0.13 mJy [\opentriangle], $r$ - 0.25 mJy [\opensquare], $i$ - 
0.34 mJy [\fullcircle] and $z$ - 0.34 mJy [\fullsquare] bands. This multiwavelength monitoring 
campaign was planned after an optical alert in 2009 (see main text).}
\end{figure}

As a result of the cosmic expansion, observed wavelengths ($\lambda_{\rm o}$) are longer than 
emission wavelengths at the AGN ($\lambda$). Hence, our NUV-optical [3] and NIR light curves of 
Q0957+561B in 2010 correspond to UV continuum sources at $z$ = 1.41 (emission at $\lambda$ = 
1438--3706 \AA). In our recent paper [3], we have only analysed the best data from each 
telescope. This approach precluded the use of NIR data, since the photometric accuracy and time 
coverage in the $iz$ bands are worse than those in the $gr$ bands. Here, using a $\chi^2$ 
minimization (e.g., [4]), the NIR light curves are compared to the $g$-band brightness record 
($\lambda$ = 1944 \AA). We thus estimate two additional relative delays for emissions at 
$\lambda$ = 3104 \AA\ ($i$-band) and $\lambda$ = 3706 \AA\ ($z$-band): $\tau$(3104 \AA) - 
$\tau$(1944 \AA) = 9.0$^{+1.3}_{-2.1}$ days and $\tau$(3706 \AA) - $\tau$(1944 \AA) = 
14.0$^{+3.1}_{-5.3}$ days (1$\sigma$ intervals), which are used to check the inner accretion 
onto the supermassive black hole. These new delays confirm the scenario derived from the 
previous interband delays [3], i.e., all measured delays are consistent with observer-frame lags 
$\tau(\lambda) \propto \lambda^{4/3}$ between a central irradiating source and standard disk 
rings emitting at different wavelengths $\lambda$ (e.g., [6]). Interestingly, the normalization 
based on NUV-optical data, $\tau$(1944 \AA) = 9 days, also works at $\lambda$ = 3000--4000 \AA\ 
(see Fig. 2).

\begin{figure}[h]
\begin{minipage}{20pc}
\includegraphics[angle=-90,width=20pc]{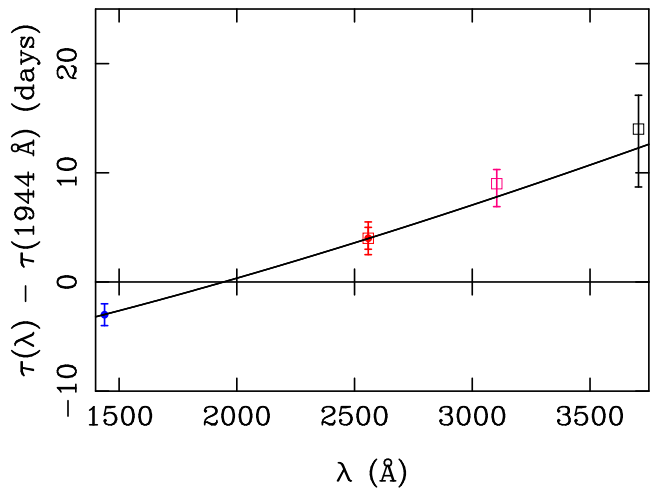}
\end{minipage}
\hspace{2pc}
\begin{minipage}{14pc}\caption{\label{Fig2}Relative delays using correlation functions 
[\fullcircle] and a $\chi^2$ minimization [\opensquare]. We show that both techniques 
(correlation functions and $\chi^2$) produce similar delays at $\lambda$ = 2558 \AA\ ($r$-band).
We also display the predictions of the centrally irradiated accretion disk with $\tau$(1944 \AA)  
= 9 days [\full].}
\end{minipage}
\end{figure}

\begin{figure}[h]
\begin{minipage}{20pc}
\includegraphics[angle=-90,width=20pc]{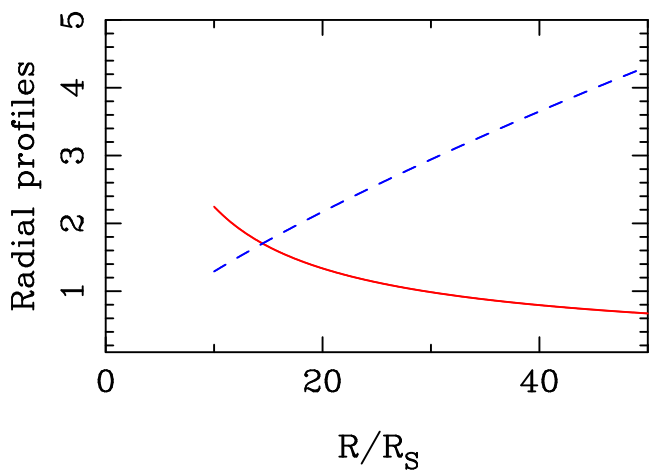}
\end{minipage}
\hspace{2pc}
\begin{minipage}{14pc}\caption{\label{Fig3}Radial profiles for the irradiated accretion disk in 
Q0957+561: temperature in 10$^4$ K [\full] and emission peak wavelength in 10$^3$ \AA\ 
[\broken].}
\end{minipage}
\end{figure}

The UV continuum sources ($\lambda$ = 1438--3706 \AA) lie within 10--30 $R_{\rm S}$ of the 
central black hole, where $R_{\rm S}$ is the Schwarzschild radius for a black hole mass $M = 2.5 
\times 10^9 M_{\odot}$ [7]. If the irradiating source is just above the black hole, the disk 
temperature at radius $R >> R_{\rm S}$ is given by (e.g., [6]) 
\begin{equation} 
T(R) = [3GM\dot{M}(1 + \alpha_{\rm iv})/8\pi\sigma R^3]^{1/4} ,
\end{equation}
where
\begin{equation}
\alpha_{\rm iv} = \frac{2(1-A)LH}{3GM\dot{M}}   .
\end{equation}
In Eqs. (1) and (2), $G$ is the gravitation constant, $\sigma$ is the Stefan constant, $\dot{M}$ 
is the mass accretion rate, $\alpha_{\rm iv}$ is the irradiation-to-viscosity ratio, $A$ is the 
disk albedo, and $L$ and $H$ are the luminosity and height of the central irradiating source. 
Each central flare propagates out at the speed of light $c$ and arrives at radius $R$ after an 
observer-frame time $\tau = (1 + z)R/c$. The temperature then rises and the peak emissivity 
increases, so each flare is reprocessed to roughly produce a fluctuation at $\lambda = 0.26 
\left[hc/kT(R)\right]$ (black-body emission peak), $k$ and $h$ being the Boltzmann constant and 
the Planck constant, respectively [6]. The expected $\tau - \lambda$ relationship is 
\begin{equation}
\tau(\lambda) = 6\ (1 + z)\left[\frac{3GM\dot{M}(1 + \alpha_{\rm iv})}{8\pi\sigma 
c^3}\right]^{1/3} \left(\frac{k \lambda}{hc}\right)^{4/3} ,
\end{equation}
and we compare this Eq. (3) to the observed law $\tau(\lambda) = (9\ {\rm days}) \times 
(\lambda/1944\ {\rm \AA})^{4/3}$. We obtain an effective mass accretion rate $\dot{M}(1 + 
\alpha_{\rm iv})$ = 3 $M_{\odot}$ yr$^{-1}$ that translates into two extreme 
accretion-irradiation regimes: 
\begin{itemize}
\nonum (a) $\dot{M}$ = 3 $M_{\odot}$ yr$^{-1}$ and $L \leq$ 10$^{45}$ erg s$^{-1}$, 
and 
\nonum (b) $\dot{M}$ = 1 $M_{\odot}$ yr$^{-1}$ and $L$ = 10$^{47}$ erg s$^{-1}$, 
\end{itemize}
assuming the reasonable constraints $4(1-A)(H/3R_{\rm S})$ = 1 and $\dot{M} \geq$ 1 $M_{\odot}$ 
yr$^{-1}$ [8]. Fig. 3 shows some radial profiles for the black hole mass and effective mass 
accretion rate of Q0957+561.

\section{Central irradiating source}
If central flares are thermally reprocessed in the disk at 10--30 $R_{\rm S}$ from the central 
supermassive black hole, what is the source of such flares? Standard simulations of X-ray 
reprocessing (e.g., [9]) ruled out the possibility that the disk variability is driven by a 
standard corona just above the black hole. The power-law X-ray emission is plausibly originated 
in the base of the Q0957+561 jet at a typical height of 200 $R_{\rm S}$ [3]. Moreover, the 
thermal source that we detect in the Chandra X-ray spectra ($kT$ = 0.08 keV) can not account for 
the observed NUV-optical-NIR variations. A central EUV (unobservable!) source seems the best
candidate to illuminate the disk and drive its variability. While standard reprocessing 
simulations rely on the lamppost model and use observed high-energy variations to try to 
reproduce observed fluctuations at lower energies (presumably originated in accretion disks; 
e.g., [9, 3] and the last paper in [2]), low-energy ($\lambda_{\rm o}$ = 3000--10000 \AA) 
brightness records and inverse problem techniques (e.g., [10]) can be powerful tools for 
reconstructing emissivities of central irradiating sources. The so-called inverse problem in 
reprocessing has a very promising future, and here we concentrate on direct reconstructions of 
the central variable (effective) luminosity of Q0957+561 based on our NUV-optical-NIR data in 
2010. 

\begin{figure}[h]
\begin{center}
\includegraphics[width=28pc]{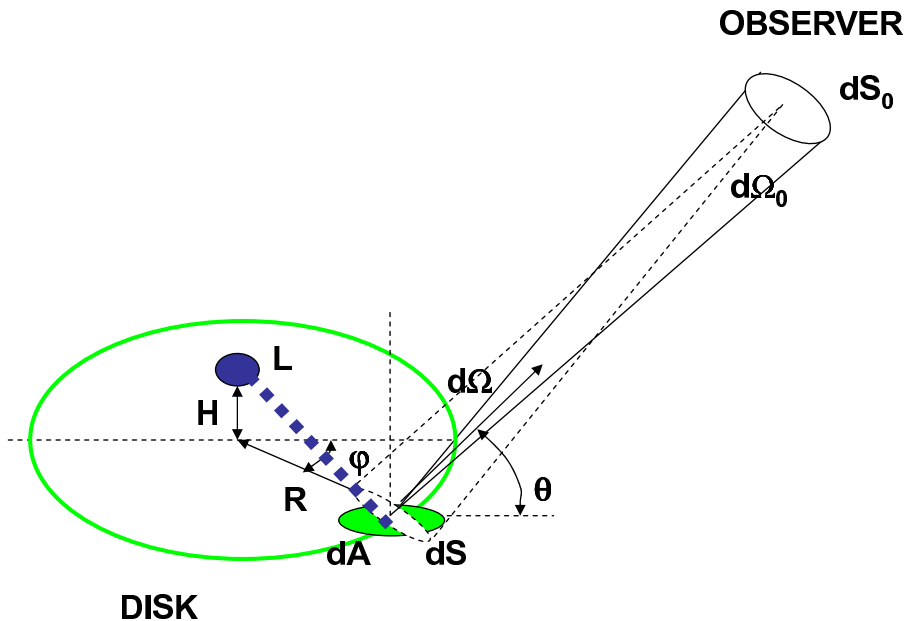}
\end{center}
\caption{\label{Fig4}The standard accretion disk is illuminated by an isotropic source of 
luminosity $L$ that is located at a height $H$ above the central supermassive black hole 
(lamppost model). A distant observer is situated at latitude $\theta$.}
\end{figure}

A fraction $1-A$ of the illuminating EUV radiation would be absorbed in the disk, and then 
reprocessed into thermal radiation. This variable illumination would cause the variable emission 
of the disk. Considering the lamppost model and an observer at latitude $\theta$ (see Fig. 4), 
the contribution of the disk portion $dA$ at ($R,\varphi$) to the flux received at time $t_0$ by 
the distant observer is related to the luminosity of the central source at time $t - 
\tau(R,\varphi; \theta)$, where $\tau(R,\varphi; \theta) = [(R^2 + H^2)^{1/2} - R \cos \theta 
\cos \varphi + H \sin \theta]/c$. As a result, the expected flux from the whole disk 
$F(\lambda_0,t_0)$ depends on different central luminosities $L(t-\tau)$ at different lags 
$\tau$ with respect to the emission time $t$. A simple approach consists of assuming an 
effective luminosity $L_{\rm eff}$ responsible for the disk emission by irradiation, which would 
correspond to an effective lag $\tau_{\rm eff}$. For each observation epoch $t_0$, one can 
minimize the difference between the modeled and observed flux: $\vert F_{\rm obs}(\lambda_0,t_0) 
- F_{\rm mod}(L_{\rm eff}) \vert$, and thus obtain the value of $L_{\rm eff}$ for the data point 
at $t_0$. We take $M = 2.5 \times 10^9 M_{\odot}$ and $z$ = 1.41, as well as a typical latitude 
$\theta$ = 45$^{\circ}$, $H = R_{\rm S}$ and $A$ = 1/4. Note that the values of $H$ and $A$ 
verify the constraint $4(1-A)(H/3R_{\rm S})$ = 1 (see above). It is not so easy to set a 
reasonable value of the AGN-observer transmission factor $\mu = \mu_{\rm lens} \times \mu_{\rm 
dust}$. This is a combination of two different contributions: the gravitational lens 
magnification ($\mu_{\rm lens} >$ 1) and the dust extinction ($\mu_{\rm dust} <$ 1) of 
Q0957+561B. We somewhat arbitrarily assume that both effects compensate each other, so that 
$\mu$ = 1. First, we check the two extreme mass accretion rates that we find from the time delay 
analysis in Section 2, i.e., (a) $\dot{M}$ = 3 $M_{\odot}$ yr$^{-1}$ and (b) $\dot{M}$ = 1 
$M_{\odot}$ yr$^{-1}$. Both extreme values lead to luminosities exceeding 10$^{47}$ erg 
s$^{-1}$, in reasonable agreement with expected emissivities in the accretion-irradiation regime 
(b). Second, we select the self-consistent regime (b), and set $\dot{M}$ = 1 $M_{\odot}$ 
yr$^{-1}$. 

\begin{figure}[h]
\begin{center}
\includegraphics[angle=-90,width=28pc]{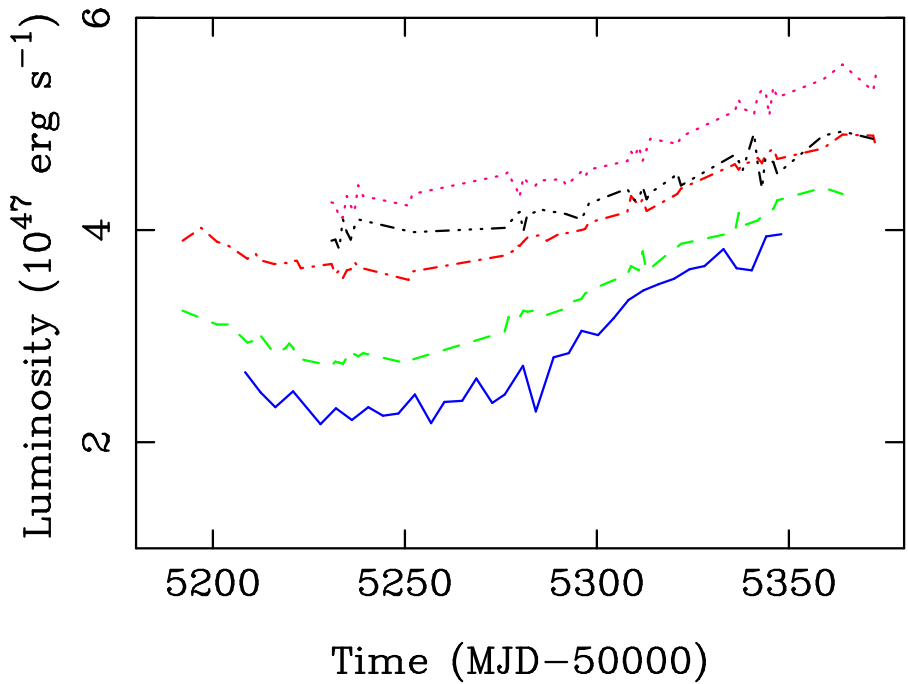}
\end{center}
\caption{\label{Fig5}$L_{\rm eff} - t_0$ laws from data in the $U$ [\full], $g$ [\broken], $r$ 
[\chain], $i$ [\dotted] and $z$ [\dashddot] bands.}
\end{figure}

We infer effective luminosities of a few 10$^{47}$ erg s$^{-1}$ from the NUV-optical-NIR records
(see Fig. 5). The effective luminosity curves are smoothed versions of the actual luminosity 
curve. Thus, the variability from the Swift/UVOT data ($U$ band) is more realistic than those 
from the optical-NIR fluxes, since the $U$-band source is relatively small and the smoothing is 
not so important. We also remark that the luminosity offsets between the $Ugri$-based  
reconstructions are plausibly due to wavelength-dependent dust extinction in the host galaxy, 
the lensing galaxy and the Milky Way. This produces a chromatic $\mu$, which has not been taken 
into account in our calculations. In addition, the ordering of our reconstructions is the 
expected arrangement for an unrealistic achromatic extinction.  

\section{Conclusions}
A recent multiwavelengtht monitoring of a sharp fluctuation in the trailing image of the 
gravitationally lensed radio-loud quasar Q0957+561 allows us to dicuss the accretion physics in 
the vicinity of a distant supermassive black hole at $z$ = 1.41. There is evidence for an 
irradiated (standard) accretion disk around the black hole, where the disk heating is mostly 
generated from irradiation by a central source. This central irradiating source very likely 
emits EUV photons and produces luminosities $\geq$ 10$^{47}$ erg s$^{-1}$. Our data also support 
the presence of an EUV luminosity variation of about 100\% on a relatively short source-frame 
timescale of 40 days. 

\ack
The authors thank Harvey Tananbaum and Neil Gehrels for granting Director's Discretionary Time 
for the Chandra and Swift observations, respectively. We also thank the operations group of the 
Liverpool Telescope (Robert Smith and Jon Marchant) for their kind interaction and support 
during our ground-based monitoring projects. The Liverpool Telescope is operated on the island 
of La Palma by Liverpool John Moores University in the Spanish Observatorio del Roque de los 
Muchachos of the Instituto de Astrof\'{\i}sica de Canarias with financial support from the UK 
Science and Technology Facilities Council. This research has been supported by the Spanish 
Department of Science and Innovation grants ESP2006-13608-C02-01 and AYA2010-21741-C03-03 
(Gravitational LENses and DArk MAtter - GLENDAMA project), and University of Cantabria funds.

\end{document}